# Tunable Two-Dimensional Electron Gas at the Interfaces of Ferroelectric Potassium Tantalate Niobates

AUTHOR NAMES

*Jiaxin Lv, [1],\* Silan Li, [1],\* Chenhao Duan,[1] Shuanhu Wang,[1] Hong Yan,[1] and Kexin Jin[1],†*

AUTHOR ADDRESS

1. Shaanxi Key Laboratory of Condensed Matter Structures and Properties and MOE Key Laboratory of Materials Physics and Chemistry under Extraordinary Conditions, School of Physical Science and Technology, Northwestern Polytechnical University, Xi'an 710072, China

ABSTRACT

The heterointerfaces at complex oxides have emerged as a promising platform for discovering novel physical phenomena and advancing integrated sensing, storage, and computing technologies. Nevertheless, achieving precise control over a two-dimensional electron gas (2DEG) in a ferroelectric oxide-based field-effect transistor (FET) configuration remains challenging. Here, we firstly demonstrate a tunable 2DEG system fabricated by depositing an amorphous $LaAlO_3$(LAO) film onto a (001)-oriented ferroelectric potassium tantalate niobate substrate. Interfaces grown under high-temperature and high-oxygen-pressure conditions exhibit a good metallic conduction. Notably, well-defined metallic 2DEGs displaying pronounced hysteresis and persistent electric-field-modulated resistance are observed below 108 K, achieving a resistance modulation of 11.6% at 7 K. These results underscore the potential for extending such behavior to other oxide-based 2DEG systems and facilitate further exploration of ferroelectric metals in complex oxide heterostructures.

## INTRODUCTION

The discovery of 2DEGs at complex oxide interfaces has opened new avenues for investigating emergent quantum phenomena and developing advanced electronic devices. Since the seminal report by Hwang and coworkers on LAO/SrTiO$_3$ heterostructures [1], extensive studies have unveiled a rich spectrum of properties, including giant persistent photoconductivity [2], two-dimensional superconductivity [3,4], a two-dimensional ferroelectric effect [5], interfacial magnetism [6], Rashba spin–orbit coupling (SOC) [7], and distinctive quantum transport behaviors [8]. Although the majority of research has concentrated on SrTiO$_3$ (STO)-based interfaces—such as those in CaZrO$_3$/STO [9], BiFeO$_3$/STO [10], GdAlO$_3$/STO [11], and PrAlO$_3$/STO [12]—recent work has broadened to explore 2DEGs in other oxide systems. Notably, Zou et al. reported a 2DEG at the LaTiO$_3$/KTaO$_3$ (LTO/KTO) interface in 2015 [13], and Zhang *et al.* demonstrated a high-mobility 2DEG at an amorphous LAO/KTaO$_3$ interface in 2017 [14]. KTaO$_3$-based 2DEGs have attracted considerable interest owing to their strong spin-orbit coupling (approximately 20 times that of STO [15,16]) and exceptional properties, including elevated superconducting transition temperatures [17–20]. Despite these advances, the coexistence of ferroelectricity and metallic conductivity at oxide interfaces remains a fundamental challenge, particularly for applications in integrated sensing, memory, and computing systems. This is largely due to the screening of long-range Coulomb interactions by itinerant charge carriers, which typically suppresses ferroelectric ordering [21]. Previous efforts to modulate resistance via ferroelectric control—such as using ferroelectric overlayers like Pb(Zr$_x$Ti$_{1-x}$)O$_3$ or BaTiO$_3$ on LAO/STO heterostructures—have achieved significant resistive switching [22–26]. More recently, Sr$_{0.99}$Ca$_{0.01}$TiO$_3$-based 2DEGs exhibiting gate-tunable transport and nonvolatile resistance states have renewed interest in ferroelectric metals [27]. However, the low Curie temperature (~35 K) of Sr$_{0.99}$Ca$_{0.01}$TiO$_3$ constrains its practical use at higher temperatures. In this work, we employ single-crystal $KTa_{0.9}Nb_{0.1}O_3$ (KTN)—a ferroelectric oxide from the 5d transition-metal family—as a substrate to host a high-quality 2DEG via deposition of amorphous LAO films. KTN undergoes a ferroelectric phase transition near its Curie temperature [28–32], offering a promising platform for coupling ferroelectricity with electronic transport. We present, for the first time, direct evidence of the coexistence of ferroelectricity and reversible resistance modulation at LAO/KTN heterointerfaces, providing a pathway toward functional oxide devices operating beyond cryogenic temperatures.

RESULTS and DISCUSSION

The experimental results underscore the distinctive properties of the LAO/KTN heterointerface, highlighting its potential as a model ferroelectric 2DEG system. As depicted in Figure 1(a), the LAO film deposited at a growth temperature ($T_g$) of 600 °C and an oxygen partial pressure [$P(O_2)$] of $8\times10^{-4}$ Pa exhibits an atomically smooth surface, with a root-mean-square (RMS) roughness of approximately 244 pm over a $1 \times 1$ $\mu m^2$ area. Figure 1(b) further reveals the well-defined interface structure. XRD pattern of the sample (Figure S1, Supporting Information) shows only Bragg reflections corresponding to the KTN substrate, confirming the amorphous nature of the deposited LAO layer. This conclusion is further corroborated by cross-sectional TEM, as shown in Figure 1(b). The interface exhibits atomic-level sharpness and flatness. As anticipated, the amorphous LAO (Area 1) is situated directly above the crystalline KTN substrate (Area 2). The structural distinction is further confirmed by FFT patterns: Area 1 displays a diffuse ring pattern without discernible diffraction spots, whereas Area 2 exhibits well-defined lattice reflections. Figure 1(c) presents a HAADF scanning TEM image and EDS mapping across the LAO/KTN interface. With the exception of Nb, all elements exhibit clearly defined spatial distributions at the interface. To determine whether the interface is 2D conductive, the out-of-plane and in-plane magnetoresistances are measured as shown in Figure S2(Supporting Information). The anisotropic magnetoresistance measurements confirm 2D conductivity at the interface.

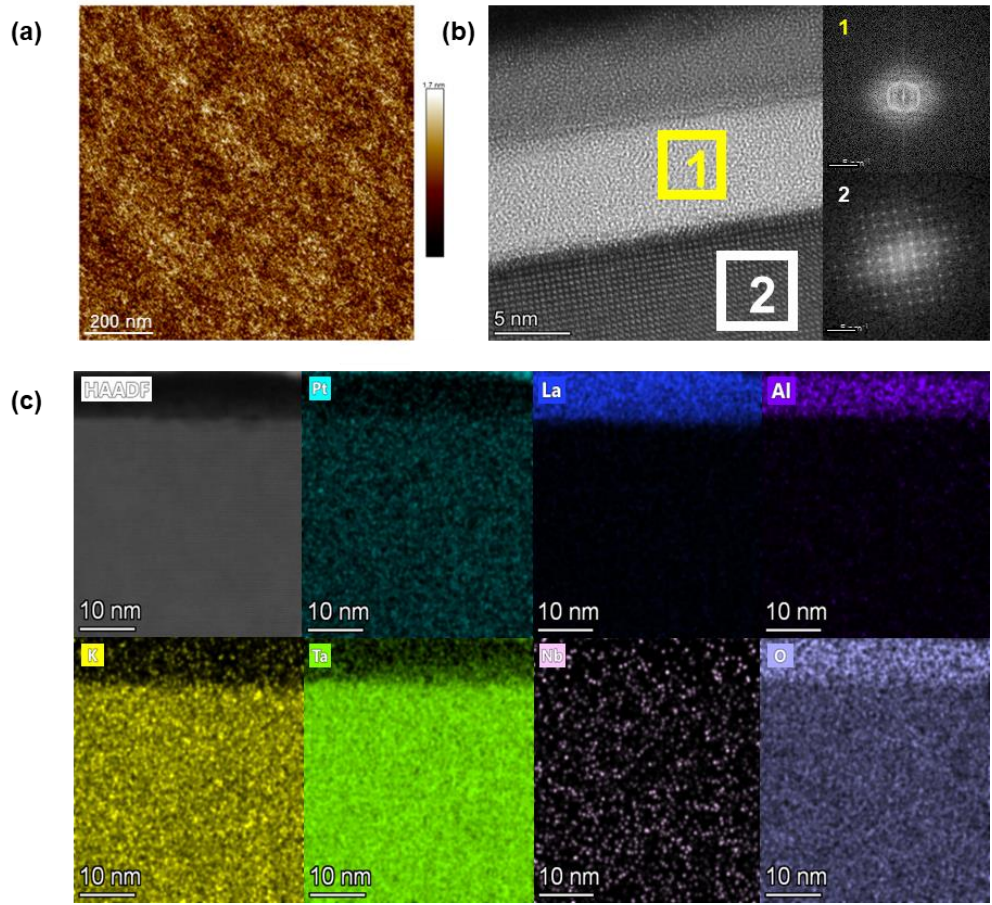

Figure. 1 Surface morphology, interface structure of LAO/KTN sample with a $T_g$ of 600 °C and an $P(O_2)$ of $8 \times 10^{-4}$ Pa and transport behaviors of different samples. (a) AFM image of the LAO film. (b) HRTEM images of the LAO/KTN interface. The insets are Fast Fourier Transform (FFT) pattern of corresponding region. (c) HAADF image and EDS mapping of different elements near the interface.

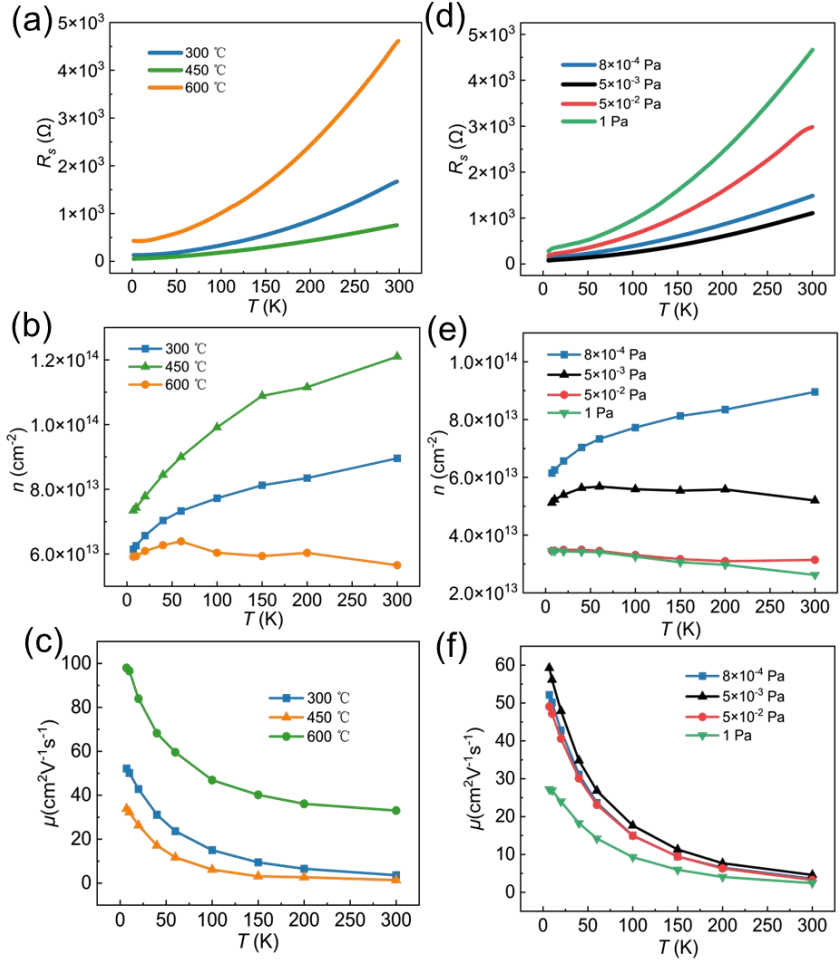

Figure. 2 Temperature dependence of the sheet resistance (a), carrier density (c) and carrier mobility (e) at a fixed oxygen pressure of $P(O_2) = 8\times10^{-4}$ Pa. Temperature dependence of the sheet resistance (b), carrier density (d) and carrier mobility (f) under different oxygen pressures at a fixed growth temperature of $T_g = 300$ °C.

We next examine the transport properties of LAO/KTN interfaces fabricated at different $T_g$ under a constant oxygen partial pressure of $P(O_2) = 8\times10^{-4}$ Pa. Figure 2(a) displays the temperature-dependent sheet resistance $R_s$ for samples prepared at various $T_g$. All samples exhibit metallic behavior, characterized by a decrease in $R_s$ with lowering temperature. The carrier density ($n$) as a function of temperature is shown in Figure 2(b). For samples grown at $T_g = 300$ and 450 °C, $n$ gradually decreases upon cooling, suggesting the presence of more defects in 2DEGs formed at lower $T_g$, which leads to carrier freeze-out at low temperatures [12]. In contrast, for the sample grown at $T_g$ = 600 °C, the carrier density not only decreases in magnitude but also becomes nearly temperature-independent, indicating a reduction in defect-mediated localization. To further probe the role of oxygen environment, we fixed $T_g$ at 300 °C and varied the oxygen pressure during growth, including $P(O_2)$ = 8×10$^{-4}$, 5×10$^{-3}$, 5×10$^{-2}$, 1 and 5 Pa. The sample grown at 5 Pa was completely insulating (not shown). As illustrated in Figure 3(d), the $R_s$–$T$ behavior of the remaining samples follows a consistent metallic trend. Oxygen vacancies play a critical role in the formation of the 2DEG in KTN-based systems. When $P(O_2)$ is increased to 5×10$^{-3}$ Pa, the carrier density (Figure 3(e)) remains relatively stable across temperatures, indicating suppression of charge localization. Further increasing $P(O_2)$ leads to a systematic reduction in n. Notably, the interface retains metallic conduction even at $P(O_2)$ = 1 Pa, exceeding the stability reported for LAO/KTO heterointerfaces [33]. The temperature-dependent Hall mobility for all samples is summarized in Figure 3(c) and (f). In all cases, the mobility increases with decreasing temperature. The highest mobility is observed in the sample grown at $T_g$ = 600 °C, reaching ∼ 98.0 cm²/V·s at 7 K. Taken together, these results demonstrate that interfaces fabricated under high growth temperature and high oxygen pressure exhibit optimal metallic characteristics with reduced disorder and enhanced carrier mobility.

Figure 3(a) presents a schematic illustration of the field-effect device configuration, in which a back-gate voltage is applied between the conductive LAO/KTN interface and a silver paste electrode deposited on the back side of KTN substrate. The ferroelectric nature of KTN substrate is confirmed by the temperature-dependent $P$-$E$ hysteresis loops shown in Figure 3(b), which exhibit characteristic ferroelectric switching. To further explore the modulation of 2DEG transport properties by ferroelectric polarization, we measured the change in resistance ($\Delta R$) as a function of applied electric field at various temperatures, as summarized in Figure 3(c).

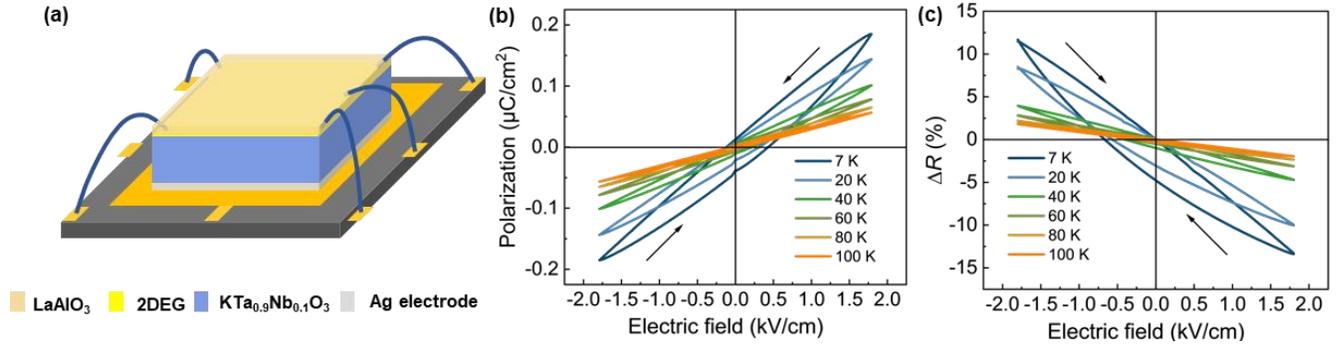

Figure. 3 (a) Schematic of the field-effect of LAO/KTN heterointerfaces. (b) *P-E* hysteresis loops measured for the sample at different temperatures ($f$ =1 kHz) (c) The $\Delta R$ of sample as a function of electric field. The arrows indicate the direction of rising and falling voltage.

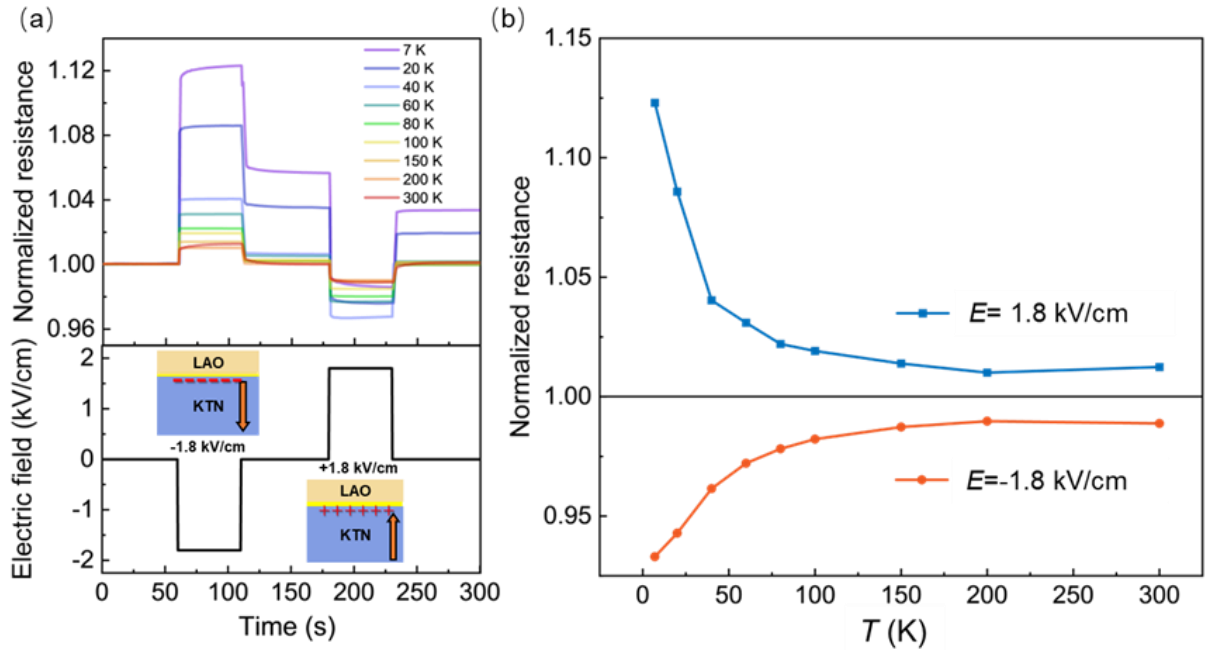

Figure. 4 (a) Time dependence of the normalized resistance of LAO/KTN heterointerface at different temperatures when the electric filed is switched at -1.8 and +1.8 kV/cm. The inserts show the changes of polarization in LAO/KTN heterointerfaces. (b) Normalized resistance of LAO/KTN heterointerface as a function of temperature at -1.8 and 1.8 kV/cm (relative to the resistance at the moment before the electric field).

Here, the persistent electric-field-induced change in the resistance before and after external electric field is defined as follows: $\Delta R = [R(E)-R(0)]/R(0) \times 100\%$, where $R(0)$ is the original resistance without the external electric field and $R(E)$ is the resistance of sample applied by an electric field. Specifically, when $T < T_c$, since the KTN is a ferroelectric single crystal, the $\Delta R$-$E$ curves of the sample have a hysteresis effect. And a lower temperature tends to increase the hysteresis effect, which is manifested as the $\Delta R$-$E$ curve being fuller and the value of $\Delta R$ after the electric field reaches the peak is larger (the maximum $\Delta R$ reaches 11.6% at 7 K). At $T > T_c$, the hysteresis of $\Delta R$-$E$ curves basically disappears. Figure 4 illustrates the time dependence of normalized resistance at different temperatures after applying -1.8 and +1.8 kV/cm. Initially, the sample stays in a state without an electric field for 60 s as a comparison. Subsequently, an electric field $E$ that switches between -1.8 and +1.8 kV/cm (the corresponding back-gate voltages are -90 and +90 V, respectively) is applied, with each field sustained for 50 s followed by a 70 s interval. Two remarkable features can be identified. One is that the resistance at $E$ of -1.8 kV/cm suddenly increases within milliseconds, and then the resistance slowly remains almost unchanged. In contrast, the resistance at $E$ of +1.8 kV/cm drops sharply in an instant due to the upward flipping of polarization, indicative of a tunable 2DEG. When a positive electric field is applied to the back gate, the resistance decreases due to the stronger ferroelectric polarization effect. Another is that after the external electric field is removed, if the temperature is higher than 108 K and the KTN is in the paraelectric phase, the resistance of sample can completely return to the initial state. If the temperature is lower than 108 K and the KTN is in the ferroelectric phase, the resistance of sample cannot completely return to the initial state. As shown in the inset of Figure 4(a), that's due to the residual polarization effect at low temperatures. The temperature dependence of the normalized resistance is presented in Figure 4(b), showing a nonlinear decrease with increasing temperature. These results collectively demonstrate that the 2DEG at the LAO/KTN interface exhibits clear ferroelectric-coupled transport features, including a kink in resistance near $T_c$ and hysteresis in the $R_s$. Notably, this system constitutes a pristine ferroelectric 2DEG, as the conducting channel is embedded within the ferroelectric host material KTN itself.

CONCLUSIONS

We have successfully realized a high-quality 2DEG with ferroelectric properties by depositing amorphous LAO films onto (001)-oriented ferroelectric KTN substrates. Our experiments reveal that interfaces formed under conditions of elevated temperature and oxygen pressure exhibit pronounced metallic characteristics, even though the LAO layer remains amorphous. Furthermore, we observe that this 2DEG exhibits tunable transport characteristics and distinct resistance states under remanent conditions. The sheet resistance shows strong modulatability through cyclic electric field application, mirroring the *P-E* hysteresis. The sheet resistance is found to be strongly tunable through cyclic electric field application, reflecting the hysteresis inherent in the polarization-electric field relationship. These findings establish a promising approach for manipulating interfacial transport via back-gate electric fields, opening new pathways for the exploration of ferroelectric metals in complex oxide heterostructures and potential applications in next-generation electronic devices.

METHODS

The 2DEGs were fabricated by depositing LAO films onto (001)-oriented KTN single-crystal substrates (5 × 5 × 0.5 mm³) using pulsed laser deposition (PLD). A systematic investigation of the transport characteristics was conducted by varying the growth temperature $T_g$ (300, 450, 600 °C) and oxygen pressure $P(O_2)$, while maintaining a constant repetition rate (1 Hz) and fluence ($\sim$1.6 J cm$^{-2}$) of the KrF excimer laser ($\lambda$ = 248 nm). The varied fabrication conditions are listed in Table S1(Supporting Information) specifically. Additionally, the dielectric constant versus temperature curve is shown in Figure. S3(Supporting Information). It is evident that the ferroelectric polarization disappears at $T_c$ = 108 K. The *P-E* hysteresis loops of KTN substrate at 80 K are shown in Figure. S4(Supporting Information) which is an evident of typical ferroelectric characteristics. The surface morphology was analyzed using atomic force microscopy (AFM; MFP-3D). Room-temperature structural characterization was performed via X-ray diffraction (XRD; Bruker D2). High-resolution transmission electron microscopy (HRTEM) and energy-dispersive spectroscopy (EDS), conducted using an FEI Themis Z instrument, were employed to assess the interfacial microstructure and elemental distribution. Transport measurements were carried out in Van der Pauw geometry using a physical property measurement system (PPMS; CFMS-14T).

ASSOCIATED CONTENT

(Word Style "TE_Supporting_Information"). **Supporting Information**. A listing of the contents of each file supplied as Supporting Information should be included. For instructions on what should be included in the Supporting Information as well as how to prepare this material for publications, refer to the journal's Instructions for Authors.

Material preparation, XRD patterns, anisotropic magnetoresistance measurements, the dielectric constant versus temperature curve and the *P-E* hysteresis loops.

AUTHOR INFORMATION

**Corresponding Author**

**\*Kexin Jin - Shaanxi Key Laboratory of Condensed Matter Structures and Properties and MOE Key Laboratory of Materials Physics and Chemistry under Extraordinary Conditions, School of Physical Science and Technology, Northwestern Polytechnical University, Xi'an, 710072, China; \*E-mail: jinkx@nwpu.edu.cn**

† Corresponding author: jinkx@nwpu.edu.cn

**Author Contributions**

(Jiaxin Lv and Silan Li) # These authors contributed equally. The manuscript was written through contributions of all authors. All authors have given approval to the final version of the manuscript.

**Funding Sources**

This work is supported by the National Natural Science Foundation of China (Nos. 52373253 and 12374192) and the Natural Science Basic Research Program of Shaanxi (Program No. 2025SYS-SYSZD-037, 2025JC-YBQN-017).

REFERENCES

(Word Style "TF_References_Section"). References are placed at the end of the manuscript. Authors are responsible for the accuracy and completeness of all references. Examples of the recommended format for the various reference types can be found at

http://pubs.acs.org/page/4authors/index.html. Detailed information on reference style can be found in *The ACS Style Guide,* available from Oxford Press.